\documentclass[twocolumn,prl,aps]{revtex4-1}
\usepackage{amsfonts}
\usepackage{amsmath}
\usepackage{amssymb}
\usepackage{color}
\usepackage{graphicx}

\begin{document}

\title{Non-Abelian operations on Majorana fermions via single charge control}
\author{Karsten Flensberg}
\affiliation{Department of Physics, Harvard University,
Cambridge, Massachusetts 02138, USA\\and Nano-Science Center \&
Niels Bohr Institute, University of Copenhagen,
Universitetsparken 5, DK-2100 Copenhagen, Denmark}
\date{\today}
\begin{abstract}
We demonstrate that non-Abelian rotations within the degenerate
groundstate manifold of a set of Majorana fermions can be
realized by the addition or removal of single electrons, and
propose an implementation using Coulomb blockaded quantum dots.
The exchange of electrons generates rotations similar to
braiding, though not in real space. Unlike braiding operations,
rotations by a continuum of angles are possible, while still
being partially robust against perturbations. The quantum dots
can also be used for readout of the state of the Majorana
system via a charge measurement.
\end{abstract}
\maketitle

Elementary excitations of topological materials can have
unusual properties, such as statistics different from that of
fermions or bosons. The most interesting possibility is
non-Abelian statistics which is believed to realized, for
example, in the 5/2 fractional quantum Hall system and in
topological superconductors, where the quasiparticles are
Majorana fermions\cite{Nayak2008}. Following the original
suggestion by Kitaev\cite{Kitaev2001}, recent theoretical
achievements have shown that topological superconductors should
be realizable in semiconductor or metallic systems with the
right combination of spin-orbit coupling, induced
superconductivity and applied magnetic
field\cite{Fu2008,Fujimoto2008,Sato2009,Lee2009,Oreg2010,Alicea2010,Lutchyn2010,Sau2010,Potter2010,QiZhang2010,Lutchyn2010b},
which has brought new optimism into the search. Several methods
to observe the non-Abelian nature of particle exchange have
been suggested. For fractional quantum systems, interferometry
of paths\cite{DasSarma2005,inter} and Coulomb blockade peak
spacings\cite{FQHCB} sensitive to the number of enclosed
quasiparticles have been analyzed, and the former recently
investigated experimentally\cite{Willett2010}. The non-local
nature combined with Coulomb interactions has also been
exploited theoretically\cite{Fu2010}. For semiconductor wire
systems, exchange of particles in one-dimensional network
structures was recently proposed as a direct way to observe
non-Abelian features of Majorana bound
states\cite{Alicea2010b}.

Moreover, non-Abelian quasiparticles have been suggested as a
basis for topological quantum computing\cite{Kitaev2003}, where
a computational step is done by physically exchanging the
positions (braiding) of the quasiparticles, thus performing a
unitary rotation in the degenerate groundstate mani\-fold. In
principle, the exchange operation only depends on the topology
of the exchange and the system is therefore argued to be robust
against local perturbations\cite{Kitaev2001,DasSarma2005}.
However, braiding of Majorana fermions has to be supplemented
with gates that are not topologically protected in order to
give a set of universal operations. Several methods have been
put forward, including bringing the quasiparticles together for
a certain amount of time\cite{Bravyi2006}, and very recent
proposals combine topological qubits and conventional
qubits\cite{Hassler2010,Jiang2010,Bonderson2010} for
implementation of quantum gates.

In this Letter, a different method for performing operations on
a set of Majorana bound states is analyzed. The individual
operations are adiabatic tunnel processes of a single electron
from a Coulomb blockaded quantum dot coupled to one or two
Majorana modes. The processes result in rotations of the
groundstate manifold, reminiscent of actual physical exchange
of the particles. Like real-space braidings, the tunnel-braid
operations are robust against dephasing. However, they are
sensitive to electrical noise on the tunnel amplitudes, and in
this respect not topologically protected. The electrical
control, on the other hand, allows rotations with tunable
angle. Moreover, to achieve the maximal protection of the
tunnel braids, a certain phase difference (controllable by a
flux) between the two Majorana bound states is needed.

The system we have in mind is a network of semiconducting wires
or a two-dimensional electron gas with spin-orbit coupling,
induced superconductivity and applied magnetic field, and in
addition a number of quantum dots, see the sketch in
Fig.~\ref{fig:manydots}(a). For parameters where the
superconductor is in the topological phase, Majorana bound
states (MBS) exist at the ends of the wires
\cite{Oreg2010,Alicea2010,Lutchyn2010,Sau2010}. They are zero
energy solutions to the Boguliubov-de Gennes equations and have
the general form
\begin{equation}\label{gammadef}
    \gamma_i=\int\! d{\mathbf{r}}\! \left(f_i
    \Psi_\uparrow^{{}}+f_i^*\Psi_\uparrow^\dagger
+g_i\Psi_\downarrow^{{}}+g_i^*\Psi_\downarrow^\dagger\right).
\end{equation}
The MBS operators obey $\gamma_i=\gamma_i^\dagger$ and
$\gamma^2_i=1$. Here $f_i$ and $g_i$ are functions of the
spatial coordinate $\mathbf{r}$ and
$\Psi_\sigma=\Psi_\sigma(\mathbf{r})$ is the electron field
operator. The Majorana bound states are tunnel coupled to
quantum dots in Coulomb blockade regime, i.e. the charge state
on a dot is restricted to be either $N$ or $N+1$ electrons,
with $N$ some arbitrary number. The level spacing of a dot is
supposed to be much larger than temperature, so that only one
quantum state is involved. Furthermore, because a large
magnetic field is applied to induce the topological
superconducting phase, it is assumed that spin degeneracy of
the dots is broken, so that we only have to consider a single
spin direction, say, spin up. The dot potential energies can be
controlled by electrostatic gates.
\begin{figure}[ptb]
\centering
\includegraphics[width=.4\textwidth]{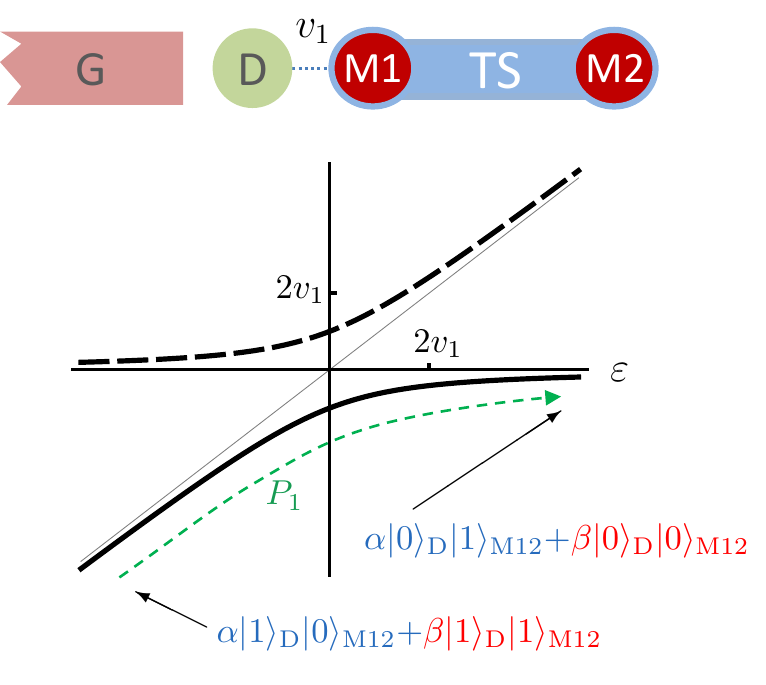}
\caption{ (Color online)
A single Majorana state (M1) coupled to a quantum dot (D) by
a tunnel coupling $v_1$. The Majorana state M1 is combined with another Majorana bound state, M2,
to form a fermion, $M12$. Starting in the groundstate, the operation $P_1$  takes the
energy $\varepsilon$ (using the gate (G)) from negative to positive values causing an
adiabatic transition from a full dot to an empty dot. The electron is added to the superconductor,
thus inverting the parity of the Majorana system. Because the total even (red) and odd (blue) parity states
are degenerate the inversion is independent on the details of the transition.\label{fig:onedot}}
\end{figure}

Projection onto Majorana bound states $\{\gamma_i\}$ coupled to
an electron level (described by a spin-up electron annihilation
operator $c$) leads to an effective low-energy tunnel
Hamiltonian
\begin{equation}\label{HTmany}
    H_T=\sum_{i}(v_{i}^{{}} c-v_{i}^* c^\dagger)\gamma_i,
\end{equation}
where $v_{i}$ is the Hamiltonian overlap between the electron
wavefunction, $\phi$, and the spin-up electron component of the
MBS $v_{i}= \langle f_i|H|\phi\rangle$. This matrix element
will be important for the actual rotation done to the Majorana
system and should therefore be controllable.

First, we consider the situation with a single MBS $\gamma_1$
coupled to a single dot and study the transition of the
groundstate as the occupancy of the dot is changed
adiabatically. Two Majorana fermions are needed to define a
usual (Dirac) fermion. Using Majorana states $\gamma_1$ and
$\gamma_2$ as basis, a fermion M12 with annihilation operator
$d=(\gamma_1+i\gamma_2)/2$ is defined. The two eigenstates of
$d^\dagger d$ are denoted $|0\rangle_{M12}$ and
$|1\rangle_{M12}$. The Hamiltonian for a dot level $c$
connected to a Majorana mode $\gamma_1=d+d^\dagger$ is then
\begin{equation}\label{H1dot}
    H_1= \varepsilon c^\dagger c+(v_1^*c^\dagger-v_1c)(d+d^\dagger),
\end{equation}
where $\varepsilon$ is the dot level energy, measured relative
to the chemical potential of the superconductor. The Hilbert
space contains 4 states, and the Hamiltonian is block diagonal
by conservation of the total parity. For both even and odd
parity the Hamiltonian matrix is
\begin{equation}\label{H2}
   H_{1,\mathrm{even/odd}}=\left(
                  \begin{array}{cc}
                    0 & v_1 \\
                    v_1^* & \varepsilon \\
                  \end{array}
                \right),
\end{equation}
with the basis for even and odd total parity cases being
$\{|0\rangle_\mathrm{D}|0\rangle_\mathrm{M12},|1\rangle_\mathrm{D}|1\rangle_\mathrm{M12}\}$
and
$\{|0\rangle_\mathrm{D}|1\rangle_\mathrm{M12},|1\rangle_\mathrm{D}|0\rangle_\mathrm{M12}\}$,
where 0(1) represent an empty(full) fermion state, and
$|\cdot\rangle_\mathrm{D}$ denotes the dot state.

Now an operation $P_1$ that adiabatically changes
$\varepsilon/v_1$ from $-\infty$  to $+\infty$ is defined, see
Fig.~\ref{fig:onedot}. If the original state of the Majorana
system is
$|i\rangle_M=\alpha|0\rangle_\mathrm{M12}+\beta|1\rangle_\mathrm{M12}$
and the dot state is $|1\rangle_\mathrm{D}$, the state of the
dot plus Majorana system is at any gate potential $\varepsilon$
given by the superposition (up to an overall dynamical phase
factor)
\begin{align}\label{psi1}
|\psi\rangle&=a(\varepsilon)|1\rangle_\mathrm{D}\big(\alpha|0\rangle_\mathrm{M12}+\beta|1\rangle_\mathrm{M12}\big)
\notag\\
&\quad +b(\varepsilon)|0\rangle_\mathrm{D}\big(\alpha|1\rangle_\mathrm{M12}+\beta|0\rangle_\mathrm{M12}\big),
\end{align}
where $v_1 b(\epsilon)=E a(\varepsilon)$ and
$E=\varepsilon/2-\sqrt{(\varepsilon/2)^2+v_1^2}$. The dot and
the Majorana system is thus entangled during the operation, but
not at the end of the operation where
$\varepsilon/v_1\rightarrow\infty$ and $b\rightarrow 1$.
Therefore, the operation $P_1$ generates an inversion of the
occupation of the M12 fermion, which in basis independent
notation\cite{basis} is expressed as
\begin{equation}\label{Pdef}
P_1:\quad   |i\rangle_\mathrm{M} \mapsto
\gamma_1|i\rangle_\mathrm{M}.
\end{equation}

If a number of Majorana bound state are connected to quantum
dots, it is thus possible to manipulate the Majorana system by
repeated applications of $P$ operations, yielding a new state
$\gamma_1\ldots\gamma_m|i\rangle_\mathrm{M}$. However, because
$\{\gamma_i,\gamma_j\}=2\delta_{ij}$ this only provides a
finite number of operations. Interestingly, two consecutive
operations $P_1P_2$ corresponds to a real-space process where
one Majorana bound state is rotated around the
other\cite{Ivanov2001}.

A natural question is how sensitive the operations $P_i$ are to
decoherence of the dot charge. To investigate this, we add
environmental degrees of freedom that couple to the dot charge.
At beginning of the $P_1$ process the initial state is
$|\Psi_i\rangle
=|1\rangle_\mathrm{D}|i\rangle_\mathrm{M}|n\rangle_\mathrm{env}$,
where $|n\rangle_\mathrm{env}$ is the initial environment
state. When expanding in powers of the tunneling Hamiltonian,
the time-evolution operation can be collected in powers of
$\gamma_1$ as $U(T,0) = U_0+\gamma_1U_1$. If the transition is
done slowly enough to guarantee that the occupancy of the dot
changes by one (can be checked by measuring the charge on the
dot) only the part $\gamma_1U_1$ survives, and Eq.~\eqref{Pdef}
holds, regardless of the dot's coupling to the environment.
\begin{figure}[ptb]
\centering
\includegraphics[width=.45\textwidth]{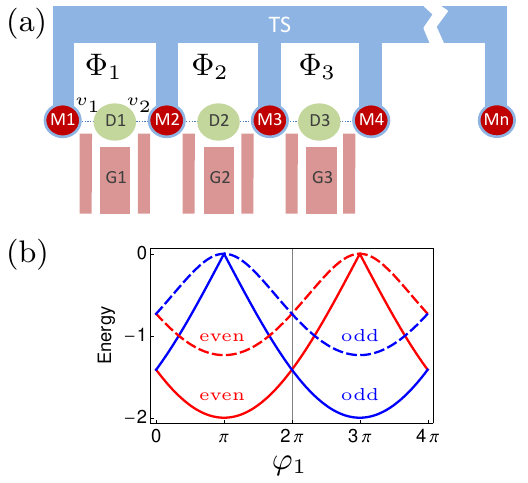}
\caption{
(Color online) (a) An one-dimensional array of Majorana states (M1,...,Mn) coupled
to quantum dots (D1,D2,...) in the Coulomb blockade regime.
Each dot is tunnel coupled to two Majorana states with
tunnel barriers (controlled by the gates adjacent to the plunger gates (G1,G2,...)).
Changing the occupancy of a dots by one electron creates the unitary rotations $P_{ij}$.
(b) The groundstate energy of one dot coupled to two Majorana
bound states, with $|v_1|=|v_2|$ for even (red) and odd (blue)
total parity of a dot and its two connecting MBS.
Even and odd cases are degenerated for $\varphi_1=2n\pi$, which makes the
$P_{ij}$ operations partially protected. The
full and dashed lines are for $\varepsilon/|v_1|=0$ and 2,
respectively.\label{fig:manydots}}
\end{figure}

As mentioned, the single MBS processes in Eq.~\eqref{Pdef} only
allow a limited set of operations. A richer set is possible if
the dots are connected to \textit{two} Majorana modes, as in
Fig.~\ref{fig:manydots}(a). Using the same basis as above, the
even and odd sector Hamiltonians for MBS $\gamma_1$ and
$\gamma_2$ coupled to a dot (D1 in Fig.~2) are in this case
\begin{align}\label{Heo}
   H_{12,\mathrm{even/odd}}&=\left(
                  \begin{array}{cc}
                    0 & v_\mathrm{even/odd} \\
                    v_\mathrm{even/odd}^* & \varepsilon \\
                  \end{array}
                \right),
\end{align}
where $v_\mathrm{even/odd} =v_1\mp iv_2$. In general, the
matrix elements $v_\mathrm{even}$ and $v_\mathrm{odd}$ are
different and the degeneracy between the even and the odd cases
is lifted. The groundstate energies are (see Fig.~2(b))
\begin{equation}\label{Eevenodd} E_\mathrm{even/odd}
=\varepsilon/2-\sqrt{(\varepsilon/2)^2
+v^2\mp2|v_1v_2|\sin(\varphi_1/2)},
\end{equation}
where $v^2=|v_1|^2+|v_2|^2$ and
$\varphi_1=2\mathrm{Arg}(v_1/v_2)$. The phase different is
controlled by the flux $\Phi_1$ and up to constant phase shift,
we can write $\varphi_1=\Phi_1/\Phi_0$, with $\Phi_1$ being the
flux in loop 1 and $\Phi_0=h/2e$.  Note that
Eq.~\eqref{Eevenodd} is 4$\pi$ periodic in
$\varphi_1$\cite{Fu2009b}.

Again, consider an adiabatic process that transfers an electron
from a dot but this time to the two Majorana states $\gamma_1$
and $\gamma_2$. Unlike the case with a single MBS, the
resulting rotation of the Majorana system is in general not
independent on the time spend in the adiabatic process, because
of the energy difference in Eq.~\eqref{Eevenodd}, see
Fig.~\ref{fig:manydots}(b). The degeneracy is restored only
when the phase difference is $\varphi_1=2n\pi$ ($n$ integer),
which therefore requires tuning the magnetic flux $\Phi_1$
(Fig.~\ref{fig:manydots}(a)). At this degeneracy point,
$v_1/v_2$ is real which allows the Hamiltonian to be written as
\begin{equation}\label{HTmanybc}
    H_{12}=\varepsilon \tilde{c}^\dagger \tilde{c} +v(\tilde{c}^\dagger-\tilde{c})
    \gamma_{12},
\end{equation}
where a new Majorana operator is defined
\begin{equation}\label{gamma12}
    \gamma_{12}=\frac{1}{v}  (|v_1|\gamma_1+|v_2|\gamma_2),
\end{equation}
and where a common phase is absorbed into the dot electron
operator $\tilde{c}=c\exp(i\mathrm{Arg}(v_1))$. Thus, since the
Hamiltonian \eqref{HTmanybc} has the same form as
\eqref{H1dot}, a dot coupled to two MBS reduces (at the
degeneracy point) to a dot coupled to a single Majorana state
$\gamma_{12}$. The conclusion from above therefore also carries
over: by adiabatically changing the electron number of the dot,
the following rotation is performed
\begin{equation}\label{P12def}
P_{12}:\quad   |\mathrm{i}\rangle \mapsto
\gamma_{12}|i\rangle.
\end{equation}

To understand the rotations that can be generated by repeated
applications of $P_{12}$ (with different ratios $|v_1/v_2|$),
we use the following Pauli matrixes acting on the two level
system spanned by $\gamma_1$ and $\gamma_2$:
$\sigma_x=\gamma_1,\sigma_y=\gamma_2$, and
$\sigma_z=-i\gamma_1\gamma_2$. In this language, the operation
$P_{12}$ makes a $\pi$-rotation around an axis in the $x$-$y$
plane, but other rotation angles around lines in the $x$-$y$
plane cannot be done. In contrast, when applying a pair
$P_{12}P_{12}'=(u\gamma_1+v\gamma_2)(u'\gamma_1+v'\gamma_2)=
(uu'+vv')+i(uv'-vu')\sigma_z$ a rotation around the $z$-axis
with tunable angle is performed. A braid operation also rotates
around the $z$-axis, but by an angle restricted to $\pi/2$.
Instead, using four MBS and the even-parity subspace to define
a qubit\cite{Bravyi2006}, a universal set of single qubit
rotations is in fact generated by pairs of $P$ operators.
Again, $P_{12}P_{12}'$ is a rotation around the $z$-axis (in
the basis $\{(00),(11)\}$ defined below), whereas
$P_{23}P_{23}'$ now gives a rotation around the $x$-axis with
controllable angle. However, two-qubit gates are not possible
in the latter representation using $P$-operators only, see
Refs.~\cite{Nayak2008,Bravyi2006} for more on quantum computing
implementations.

A special and illuminating case is when the dots couple with
equal strength to two MBS ($|v_{1}|=|v_{2}|$ in
Eq.~\eqref{gamma12}), which results in operators
$F_{i}=\frac{1}{\sqrt2}(\gamma_i+\gamma_{j+1})$ acting on
nearest neighbors. They are related to braid operators
$B_{i}=\frac{1}{\sqrt2}(1+\gamma_{i+1}\gamma_{i})$\cite{Ivanov2001}
by  $B_{i}=F_{i}\gamma_i=\gamma_{i+1}F_{i}$. The $F_i$
operators fulfill  $F_i^2=1$ and
\begin{subequations}
\begin{eqnarray}
  F_i F_j &=& -F_j F_i,\quad |i-j|>1\\
  F_i F_{i+1}F_i &=& -F_{i+1} F_{i}F_{i+1},
\end{eqnarray}
\end{subequations}
which differs by a minus sign from the relations defining the
braid group\cite{Nayak2008}. As a side remark, $F_i$ form a
projective representation of the permutation
group\cite{project}.

To demonstrate the non-Abelian nature of the tunnel-braid
operations, consider now an explicit example with four Majorana
states and three dots. The state of the superconductor is
initialized by tuning the dots and the magnetic field to fuse
Majorana pairs (1,2) and (3,4) and letting them relax. The
initial state is
$|00\rangle=|0\rangle_\mathrm{M12}|0\rangle_\mathrm{M34}$,
referring to the occupation of the fermions,
$d_1=(\gamma_1+i\gamma_2)/2$ and $d_2=(\gamma_3+i\gamma_4)/2$.
We will consider applications of pairs of $F_{i}$ and hence
restrict to the subspace of even parity, spanned by
$|00\rangle$ and $|11\rangle=d_2^\dagger
d_1^\dagger|00\rangle$. The possible unitary transformation are
given by
\begin{align}
  \left(F_{1} F_{2}\right)_\mathrm{even}&=\left[\left(F_{2} F_{3}\right)_\mathrm{even}\right]^T=\frac{1}{\sqrt{2}}\left(
\begin{array}{cc}
1 & -i \\
1 & i
\end{array}
\right),
\end{align}
and $\left(F_{1}F_{3}\right)_\mathrm{even}=\sigma_x$ (up to
phase factors). Other permutations can be deduced from $F_{i}
F_{j}=[F_{j} F_{i}]^{-1}$. Application of $F_{1}F_2$ or
$F_{2}F_1$ gives
\begin{subequations}\label{nonA}
\begin{align}
F_{1}F_{2}|00\rangle &= \frac{1}{\sqrt{2}}\left(|00\rangle+| 11\rangle\right),\\
F_{2}F_{1}|00\rangle &= \frac{1}{\sqrt{2}}\left(|00\rangle+i|11\rangle\right).
\end{align}
\end{subequations}
For one sequence the resulting state is an eigenstate of
$\sigma_x$ and for the other it is an eigenstate of $\sigma_y$.
However, expectation values of neither of these operators are
easily measured using quantum dots, capable of measuring charge
only. Initial application of $F_{2}F_{3}$ and $F_{3}F_{2}$ in
Eqs.~\eqref{nonA} rotates to states that can be distinguished
by an occupation measurement and this yields
\begin{subequations}
\begin{align}
F_{1}F_{2}F_{2}F_{3}|00\rangle &= |11\rangle,
\label{P12232334}\\
F_{2}F_{1}F_{3}F_{2}|00\rangle &= |00\rangle.
\end{align}
\end{subequations}
The last sequence can be implemented in the following way: the
potentials on the dots D1, D2, and D3 in
Fig.~\ref{fig:manydots}(a) are successively increased from
positive to negative voltages, thus emptying the dots, and then
D2 is filled again by tuning the D2 potential back to positive
voltages. The sequence \eqref{P12232334} is done in the same
way, except in a different order.

Finally, we discuss how the state of the Majorana system can be
read out. Several methods have been proposed to measure the
state of coupled Majorana modes, including observing changes in
the current-phase relation of a Josephson junction with two
Majorana states in the loop\cite{Fu2008,Fu2009}, or by
interferometry\cite{Nayak2008,inter}. In the present setup it
is natural to use the quantum dots, which is indeed possible.
By choosing the phase difference between two MBS, so that the
fused even or odd states are split in energy, the occupancy can
be read from the adiabatic curves similar to the one in
Fig.~\ref{fig:onedot}, because the off-diagonal elements in
Eq.~\eqref{Heo}, unlike in Eq.~\eqref{H2}, differ for the even
and odd cases. Maximal visibility is achieved for
$\varphi_1=\pi$, where the two off-diagonal matrix elements are
$|v_1|\pm|v_2|$ for even/odd, respectively.

In conclusion, a scheme for manipulation of the state of a set
Majorana fermions has been proposed. It allows for
demonstration of the non-Abelian nature of the quasiparticles,
albeit not by real-space exchanges. Instead the exchanges are
performed via removal or addition of real electrons, made
possible by Coulomb blockade.

\acknowledgments

L. Fu, B.~I. Halperin, M. Leijnse, R.~M. Lutchyn,
V.~E.~Manucharyan, and C.~M. Marcus are gratefully acknowledged
for discussions and A. Ipsen and H. Moradi for explanation of
reference\cite{project}. Research funded in part by The Danish
Council for Independent Research $|$ Natural Sciences and by
Microsoft Corporation Project Q.


\begin{thebibliography}{10}
\bibitem{basis} Note that the MBS operator $\gamma_2$ drops
    out, since it was only needed to define a basis for the M12 fermion.
\end{thebibliography}
\end{document}